# On Social and Economic Spheres:
# An Observation of the "*gantangan*" Indonesian tradition

Hokky Situngkir[1], Yanu Endar Prasetyo[2]

[1]Dept. Computational Sociology, Bandung Fe Institute (BFI)
[2]Center for Appropriate Technology Development, Indonesian Institute of Sciences (LIPI)

**ABSTRACT**

Indonesian traditional villagers have a tradition for the sake of their own social and economic security named "*nyumbang*". There are wide variations of the traditions across the archipelago, and we revisit an observation to one in Subang, West Java, Indonesia. The paper discusses and employs the evolutionary game theoretic insights to see the process of "*gantangan*", of the intertwining social cohesion and economic expectation of the participation within the traditional activities. The current development of the "*gantangan*" tradition is approached and generalized to propose a view between the economic and social sphere surrounding modern people. While some explanations due to the current development of "*gantangan*" is drawn, some aspects related to traditional views complying the modern life with social and economic expectations is outlined.

## 1. INTRODUCTION

We are all living in two spheres: economic and social. The modern life of capitalism has delivered us to the existing that made an obvious borderline of both spheres. While economic sphere is often related to the environmental assets, which are considered as natural capital which has both limited and fragile tendencies, social sphere is associated to any cultural forms, symbolic bonds and community infrastructures, called social capital upon which an edifice of economic performance is made [2]. The intertwining of both spheres is frequently related to the "irrationality" behaviors as discussed in some popular works of modern human and social life [*cf.* 1].

A mother serving the family with delicious meals in thanksgiving are not supposed to receive any payments from her working children, no matter how richer their children are. However, in the other hand, a small help from a hotel officer would be only favored in an expectation of small amount of money. The intertwining of social and economic sphere surrounds a modern human life. The behavior based on good understanding with both spheres is a reflection of maturity or social intelligence [4]. The intertwining social and economic spheres are even later elaborated into four spheres, including the natural system and system regulations via political organizations [9]. Our understanding of the spheres surrounding modern human life, nevertheless, would give some important insights understanding some scheme organizing the completely social system.

The modern people in a developing country like Indonesia witness the emerging form of social life from the social and economic spheres. While the traditional culture of Indonesia practices a very strong tendencies and values for sharing [*cf.* 8], especially those who live in the rural areas of villages, the value for economic expectation, as consequence of living in the age of modern capitalism, may sometimes conflict one another [*cf.* 11]. The social values for sharing of people in the villages of Indonesia are conventionally recognized as the tendency for homogenous social life. Sharing treasures is traditionally common, to reach the equal ownership of economic goods, even when all the people within the village are in poverty, a fact that is known as "shared poverty" [3].

One of traditional practices in Indonesian villages is the tradition of "*nyumbang*". *Nyumbang* is the practice of giving away some fortunes and treasures for other people as an act of helping others, of those who have better social and economic status. The tradition still exists even today; in the practices of people give some of their fortunes to other people having a celebration or feast (in Indonesia, commonly called *hajatan*) on wedding, islamic pilgrimage (hajj), and more. People will do the *nyumbang* in an expectation that others may do the similar thing when they shall organize other celebration or feast. While originally the *nyumbang* tradition is delivered genuinely with social motives and less economic expectation, the interaction of the traditional practices to the modern life would enhance the economic expectations within the practice of *nyumbang* [10]. The *nyumbang* tradition is called "*Jagong*" in Central Java, "*De' Nyande*" in Madura, "*Mbecek*" in Eastern Java, and "*Gantangan*" in West Java.

The transformation of the practice of *nyumbang* is observed to see the overlapping of economic and social sphere in general. The next section describes the tradition of *nyumbang,* with the focus of observation is the practice of the tradition in West Java (called "*gantangan*"). This section follows the discussions on how we can see the practice of "*gantangan*" as an evolutionary process in the framework of social and economic sphere.

## 2. "GANTANGAN" RECONCILES OF SOCIAL AND ECONOMIC SPHERES

*Gantangan* or sometimes called "*gintingan*", "*berasan*," or "*narik*" is a traditional practice in Subang, Western Java. The *gantangan* system has been delivered within the rice farming social life and in some previous years developed to be one of the economic exchange systems. When someone organize a feast, other people, be it relatives or neighbors come and "deposit" an amount of rice or money. The amount of rice or money is seen as a "loan" and somehow become a kind of "debt" in the perspective of the organizer. Someday, when the one who deposit the rice or money organizes other feast in any celebrations, the other will return and give the rice and money in the same amount she has received previously [10].

For example, if Mr. A gives 50 liters of rice (in traditional measurement, it is "5 *gantang*") and an amount of money USD 20,- to Mr. B in her daughter's wedding party, then it would be fair for Mr. A to expect that Mr. B gives him, at minimum, the same amount of rice and money when later, he organize an Islamic feast for his younger son's circumcision. The return from Mr. B is said to be minimum the amount of rice and money that he has received previously because he may add the amount due to his expectation to be given by Mr. A when next time he organize other "*hajatan*" (feast or celebration party). Thus, *gantangan* becomes a kind of credit saving for anyone in the community. That is the economic sphere of the *gantangan*.

Looking the *gantangan* merely in the aspects of economic expectation (economic sphere), an participant may choose to:

- ✓ $\alpha$ : depositing rice and/more money more than the average one would deposit with expectation to gain a return later (thus, using the process of *gantangan* as a kind of "investment" or saving),
- ✓ $\beta$ : depositing an average or "standard" or minimum amount of rice and/or money in a feast, in order to just keeping up with the social and community association,
- ✓ $\gamma$ : do not get along or "abstain" with the *gantangan* or the *hajatan* process.

The latest options may risk the exclusion of one to the wider aspects of social relations within families, neighborhood, or even friendship. People are free to choose whether to participate or not, and their participation would yield a payoff that can be written as a payoff matrix $A$,

|   | $\alpha$ | $\beta$ | $\gamma$ |
|---|---|---|---|
| $\alpha$ | $N(p_{ES} + m_{SS})$ | $N(\frac{p_{ES} + m_{SS}}{2})$ | $N(m_{SS})$ |
| $\beta$ | $N(\frac{p_{ES} + m_{SS}}{2})$ | $N(m_{SS})$ | $N(m_{SS})$ |
| $\gamma$ | $N(p_{ES})$ | $N(m_{SS})$ | 0 |

One playing the role $\alpha$, would provide herself with social excel ($m_{SS}$) and would provide as well the amount of rice and money ($p_{ES}$) in return for other players, where $p_{ES}, m_{SS} > 0$, while playing the strategy $\beta$ is providing merely the social and community association. However, there are some people play strategy $\gamma$, a kind of opportunistic strategy of gaining the gain without any 'investment' at all.

The variable $m_{SS}$ is related to the other aspects regardless the economic expectation in the participation to the *gantangan*. One with higher social and economic status tends to give more amount of rice or money. They also tend to organize more celebration party or feast including more participants within. People trust those with higher social and economic status more, since they have the capacity to giving more in return. The poorer people tend to be excluded from the whole process of *gantangan* for their inability giving rice and money in return.

Observing the pay-off matrix, we can see that the equilibrium of the game would depend on the value of $p_{ES}$ and $m_{ES}$ included within the game. The bigger $p_{ES}$ would push the game into the stronger position and dominating player playing the strategy $\alpha$, and the bigger value of $m_{SS}$ makes the stronger position of the strategy $S$.

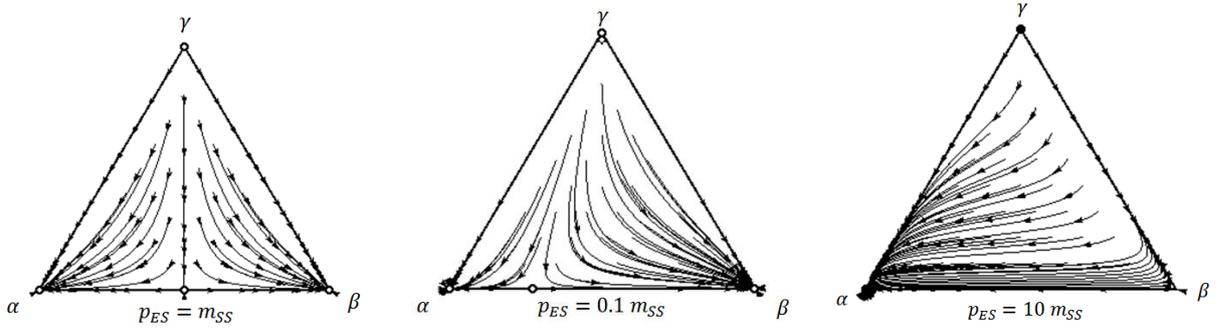

**Figure 2.** The replicator dynamics of population with economic expectation ($\alpha$), socially motivated ($\beta$), and agents choose to absence the "*gantangan*".

We denote the pay-off matrix above as $A = [a_{ij}]$, and write the deterministic replicator-mutator dynamics by denoting the frequency of strategy $i$ ($x_i$), $i = 1, 2, \ldots n$,

$$\dot{x}_i = \sum_j^n x_i f_i q_{ji} - x_i \varphi \quad (1)$$

where the fitness of strategy $i$

$$f_i = \sum_j x_j a_{ij} \quad (2)$$

and the average fitness of the whole population,

$$\varphi = \sum_i x_i f_i \quad (3)$$

and the probability of strategy $i$ having the offspring using strategy $j$, $q_{ji}$, as, $\sum_i x_i = 1$.

We can draw the replicator-mutator dynamics and find out the respective stationary state for the corresponding variations of the $p_{ES}$ and $m_{SS}$ [5, 7] as shown in figure 1. From the three strategies, we could see there are four stationary state yielded from varying the $p_{ES}$ and $m_{ES}$, and two of them are shown reflecting the purely dominated strategies of $\alpha$ and $\beta$.

An interesting fact that we can observe in here is the asymmetric $p_{ES}$ and $m_{SS}$ one another within the game as shown in the "*gantangan*" process. As the economic sphere (ES) gives larger expectation, it may strictly dominate the population, yet, the larger expectation for the gain in social sphere (SS) still slightly return the small amount of the whole population.

In fact, the asymmetric ES and SS in the development trend of the "*gantangan*" process are obviously shown empirically. There have been noted the kind of "commercialization" of the traditional "*gantangan*" as the process going along until today. The current "*gantangan*" has demonstrated to be a way people utilizing traditional culture into the way to gain resources for economic expectations. Moreover, there have been variations of "gantangan" in some particular villages where the members of the process are closed communities within the society [10]. Due to this phenomenon, the "*gantangan*" process is not exhibited from the whole population no more, but closed into several families within in groups of people in villages. Thus, in some observed villages, the reconciliation of the economic and social sphere happens to be the dominating expectations for the economic benefit, more than just the accentuation of social and cultural motives in the society. The traditional "*gantangan*" has turned into the regular social gathering whose members contribute to and take turns at gaining profit (aggregate amount of rice and sum of money) while organizing feast.

## 3. CONCLUDING REMARKS AND FURTHER WORKS

The tradition of "*gantangan*", as a specific variation of "*nyumbang*" tradition was once developed in the motivation of social and economic security among villagers in Indonesia. While people are happy with the festivals ("*hajatan*") organized, people may give some of their economic estates to the organizing family. However, the interaction of the traditional view of "*gantangan*" has been challenged by the tendency for economic expectations taught by the modern capitalistic life. The "*gantangan*" tradition may currently be viewed as an organic way of the people in Indonesian villages to invest some of their wealth with some expectations for future return. The "*gantangan*" might be a portrait of a kind of reconciliation between the intertwining economic and social sphere within the traditional society in Indonesia.

The game-theoretic model employed to analyze the social interaction in "*gantangan*" showed how such economic tendency (the recent yet dominant view of "*gantangan*" as an activity of investment) may invade the social contents of the traditional motives for the sake of social cohesion. The view seeing "*gantangan*" as a kind of investment is evolutionarily fit and invades the particularities motives of merely social cohesion. This is an opportunity to enhance economic life in the village, as well as preserving the local and traditional view on social sphere, a collective and a kind of grassroots innovation that has potential even for poverty alleviation within agricultural society.

The drawn future works related to the financial security and an insight relating to "gantangan" is probably a good start on how we eventually formalize with more technicalities the reconciliation between the social and economic sphere surrounding people in the village while embracing the modern view of the world.